\documentclass[twocolumn,showpacs,amsmath,amstex,amssymb,mathfonts,prl]{revtex4-1}
\usepackage{graphicx,bm,units}

\begin{document}

\newcommand{\vk}{{\bf k}}
\def\ns{^{\vphantom{*}}}
\def\ket#1{{|  #1 \rangle}}
\def\bra#1{{\langle #1|}}
\def\braket#1#2{{\langle #1  |   #2 \rangle}}
\def\expect#1#2#3{{\langle #1 |   #2  |  #3 \rangle}}
\def\cH{{\cal H}}
\def\half{\frac{1}{2}}
\def\sut{\textsf{SU}(2)}
\def\suto{\textsf{SU}(2)\ns_1}
\def\kF{\ket{\,{\rm F}\,}}

\title{Entanglement Spectrum of a Disordered Topological Chern Insulator}

\author{Emil Prodan$^1$, Taylor L. Hughes$^2$ and B. Andrei Bernevig$^3$}
\address{$^1$Department of Physics, Yeshiva University, New York, NY 10016}
\address{$^2$Department of Physics, University of Illinois, 1110 West Green St, Urbana IL 61801} 
\address{$^3$ Department of Physics, Princeton University, Princeton, NJ 08544} 

\begin{abstract} We investigate the behavior of a topological Chern Insulator (CI) in the presence of disorder, with a focus on its entanglement spectrum (EtS) constructed from the ground state. For systems with symmetries, the EtS was shown to contain information about the topological universality class revealed by sorting the EtS against the conserved quantum numbers. In the absence of any symmetry, we demonstrate that statistical methods such as the level statistics of the EtS can be equally insightful, allowing us to distinguish when an insulator is in a topological or trivial phase and to map the boundary between the two phases.  The phase diagram of a CI is computed as function of Fermi level ($E_F$) and disorder strength using the level statistics of the EtS and energy spectrum (EnS), together with a computation of the Chern number via an efficient real-space formula.
\end{abstract}

\pacs{63.22.-m, 87.10.-e,63.20.Pw}

\date{\today}

\maketitle

Topological insulators (TI) are materials that do not conduct electricity in the bulk but display conducting edge channels. CIs represent a particular class of TIs \cite{haldane1988} that have broken time-reversal symmetry. They haven't been observed yet experimentally, but a time-reversal invariant version has been proposed \cite{kane2005A,bernevig2006a,bernevig2006c} and discovered \cite{koenig2007}. Since then, the field of TIs became increasingly popular. The central claim of the field, and the basis for most potential applications, is the robustness of TIs' properties against imperfections. Our Letter contributes to the ongoing research \cite{Obuse2008,Nomura2007,Ryu2007,xu2006,essin2007,Onoda2007,Sheng:2006vn} on disorder effects in TIs and gives two practicle tests to determine if the ground-state of a disordered CI is in the topological or trivial phase. The tests involve only the ground-state wavefunction.

While our work concentrates on CIs, it  addresses a broader question: Given the  ground state of a Hamiltonian, how much information can we extract about its ``topological" universality class? Ref.~\cite{li2008}  suggested that the answer resides in the \emph{entanglement spectrum}, {\it i.e.\/} the full set of eigenvalues of the reduced density matrix.  For the $\nu$=$\frac{5}{2}$  Fractional Quantum Hall (FQH) states, the EtS levels and their multiplicities, when plotted versus the angular momentum,  match the levels and multiplicities of the edge modes \cite{li2008,regnault2009, ronny2010}. The EtS also captures the low-energy physics of gapless spin chains \cite{ronny2009} and for topological insulators it exhibits analogs of the physical edge state spectra \cite{haldane2009,fidkowski2010,turner2009,Kargarian2010}.

\begin{figure*}
  \includegraphics[width=17.4cm]{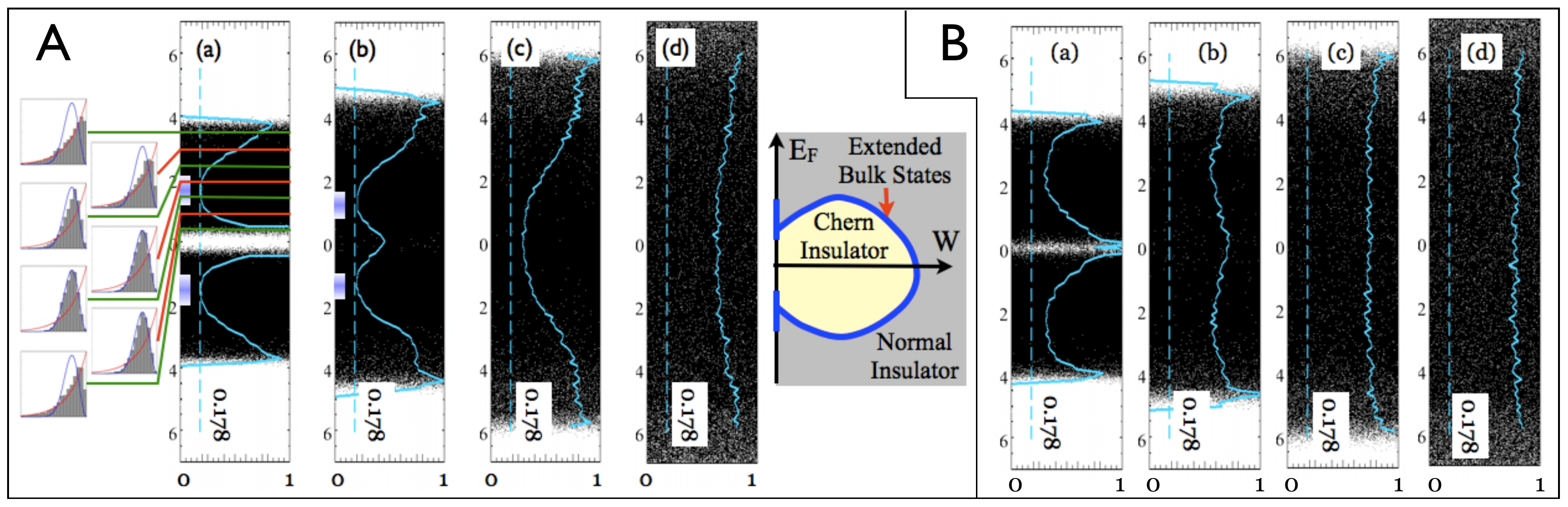}\\
  \caption{EnS and level statistics for a disordered (A) CI ($\zeta=0.3 i$) and (B) trivial insulator ($\zeta=0.3$), at disorder strengths $W=$ (a) 3, (b) 5, (c) 8, (d) 11. Panel (A) also shows the phase diagram of the disordered CI inferred from panels (a)-(d), and a few histograms of the level spacings recorded at the indicated energies. The histograms are compared with $P_{\mbox{\tiny{GUE}}}(s)$ (blue line) and $P_{\mbox{\tiny{Poisson}}}(s)$ (red lines) distributions. The blue lines overlaying the EnS are the variances of the energy-spacing distributions plotted on a scale between 0 and 1. The dashed blue line represents the variance ($\sim 0.178$) of $P_{\mbox{\tiny{GUE}}}$. }
 \label{PhaseDiagram}
\end{figure*}

Prior studies on EtS treated systems with translational invariance and the EtS of FQH states, spin chains, and topological insulators were plotted versus the momentum parallel to the cut. This is not generic and until now it was unclear if the EtS is useful when \emph{no symmetry} is present. Without translational symmetry, what remains that is fundamental? Seminal thinking by Wigner gave us the answer: adopt a statistical view of spectra. Subsequent work on random matrix theory revealed universal spectral properties that are dependent only on the fundamental symmetries of the Hamiltonians \cite{MehtaBook}. This line of thinking has had success in both many-body systems as well as in the theory of Anderson localization.  We adopt it here and apply it to the EtS rather than to the EnS. For disordered CIs, we show that the EtS gives clear signatures of whether the CI is in the topologically nontrivial or in the simple Anderson insulator state. Such signatures, present in the ground state alone, are important for developing tools to attack the interacting many-body problem with disorder since statistical analysis of the EnS is impossible for large systems - diagonalization procedures only give a few low-lying energy states. The EtS of  the ground-state, however, contains a large number of eigenvalues on which level statistics can be performed.  We compare our results for the EtS with computations of the Chern number via a real-space formula and with results from conventional level statistics of EnS.

We consider  2D lattice models with $\alpha=1,\ldots,K$ quantum states $|{\bm x},\alpha\rangle$ per each site ${\bm x}$ and Hamiltonians:
\begin{equation}\label{H0}
\begin{array}{c}
H_\omega=\sum t_{\alpha\beta}^{{\bm x}-{\bm y}} |{\bm x},\alpha\rangle \langle {\bm y},\beta|+ W\sum \omega_{{\bm x},\alpha} |{\bm x},\alpha\rangle \langle {\bm x},\alpha|,\nonumber
\end{array}
\end{equation}
where the first term is a translationally invariant insulating Hamiltonian $H_0$ and the second is a disorder potential $V_\omega$. In 2D, $H_0$ can display topological properties, manifested in the emergence of chiral edge modes. The number of  stable chiral edge modes is equal to the $C$ number of the occupied bulk states \cite{ProdanJMP2009}. By definition, a CI is an insulator with $C$$\neq$0. Its \emph{bulk} states display a spectacular behavior, manifested in the persistence of extended states even when the disorder is on. The interesting physics of the CIs is due to these states  -  the edge modes are nothing but the delocalized bulk states terminating at the edge.

In our  calculations, we use the spin-up component of the Kane-Mele Hamiltonian \cite{kane2005A} with $\lambda_R=0$ and  $\lambda_{SO}=\eta-it$ (in order to connect with previous studies \cite{Onoda2007}):
\begin{equation}\label{ModelHam}
\begin{array}{c}
H_0=\sum_{\langle \bm x \bm y\rangle } |{\bm x}\rangle \langle {\bm y}| + \sum_{ \langle\langle\bm x \bm y\rangle\rangle } \{ \zeta_{\bm x} |{\bm  x}\rangle \langle {\bm  y}|+\zeta_{\bm x}^* |{\bm  y}\rangle \langle {\bm  x}|\},\nonumber
\end{array}
\end{equation}
where $\zeta_{\bm x}=\frac{1}{2}\alpha_{\bm x}(t+i\eta)$ with $\alpha_{\bm x}$= the iso-spin of the site and ${\bm x}$, ${\bm y}$ are sites of a honeycomb lattice. This $H_0$ displays a topological phase for $|\eta| > |t|\tan \frac{\pi}{6}$.  For disorder we use uniform random entries $\omega_{ {\bm x}}\in [-\frac{1}{2},\frac{1}{2}]$.  

We first use the traditional level statistics analysis of the EnS to probe the extended/localized character of the bulk states. Fig.~\ref{PhaseDiagram} shows the EnS of $H_\omega$ at disorder strengths $W$=3, 5, 8 and 11, when $H_0$ is in topological (Fig 1A) and trivial (Fig 1B) phases. The energy levels are shown on the vertical axis for $10^3$ disorder configurations offset horizontally. Overlayed is the variance of the energy level spacings, collected at all energies using a small window.  From the histograms, we see two regions where the level-spacing distribution perfectly matches the Wigner-Dyson distribution $P_{\mbox{\tiny{GUE}}}(s)$=$\frac{32 s^2}{\pi^2}e^{-4s^2/\pi}$. The level-spacing variance $\langle s^2 \rangle$$-$$\langle s \rangle ^2$ at these energies converges to the variance of $P_{\mbox{\tiny{GUE}}}$$\simeq$0.178. We infer that these regions of level repulsion contain extended states. In the rest of the spectrum, the histograms match the Poisson distribution $P_{\mbox{\tiny{P}}}(s)$=$e^{-s}$ and the variance takes large values ($O(1)$). We infer that in these regions the states are localized. Upon increasing $W$, the regions of delocalized spectrum (where the variance is exactly $0.178$) converge towards each other to eventually collide and disappear. This is consistent with the ``levitation and annihilation" phenomenon \cite{Onoda2007} and suggests the phase diagram shown in Fig.~\ref{PhaseDiagram}A. In  contrast, if we start from the normal insulator and increase W, all the states localize (see Fig.~\ref{PhaseDiagram}B) at any disorder strength and there is no diffusive regime (a small variance in Fig.~\ref{PhaseDiagram}B(a) is a finite size effect \cite{Cuevas1998}).

\begin{figure*}
\includegraphics[width=17.4cm]{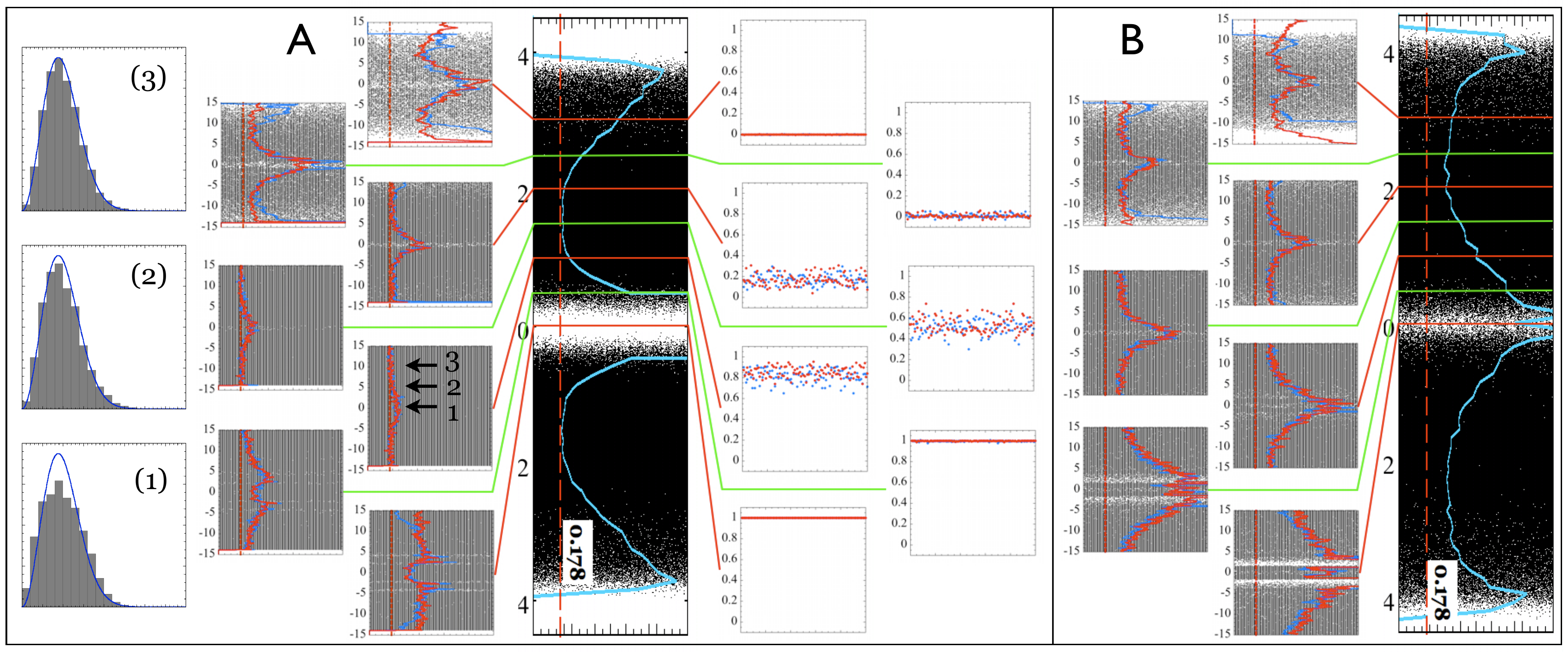}\\
\caption{(A) EtS and the variance of the level spacings (left panels), EnS (mid panel) and $C$ number (right panels) for (A) $\zeta=0.3i$ CI and (B) $\zeta=0.3$ normal insulator both with $W$=3. The EtS and $C$ were computed for seven $E_F$'s as indicated. The blue/red data correspond to calculations on 30$\times$30/40$\times$40 lattices. $C$ was identically zero for the normal insulator. The vertical dotted line in the EtS plots marks the 0.178 value. Panels (1-3) show the histograms of the level spacings collected at the entanglement energies marked by arrows. Overlayed,  is the Wigner-Dyson distribution $P_{\mbox{\tiny{GUE}}}(s)$. }
\label{chernandent}
\end{figure*}

We corroborate these results with a calculation of the Chern number $C$. For a clean system:
\begin{equation}\label{Chern1}
\begin{array}{c}
C=\frac{1}{2\pi i} \int_{\mbox{\tiny{BZ}}} \mbox{tr}\{\hat{P}({\bm k})[\partial_{k_1}\hat{P}({\bm k}),\partial_{k_2}\hat{P}({\bm k})]\}d^2 {\bm k},
\end{array}
\end{equation}
where $\hat{P}({\bm k})$ is the $k$-decomposition of the projector $P$ onto the occupied states. In real-space:
\begin{equation}\label{Chern2}
\begin{array}{c}
C=2\pi i \sum_\alpha\langle 0,\alpha|P \big{[} -i[\hat{x}_1,P],-i[\hat{x}_2,P] \big{]}|0,\alpha\rangle.
\end{array}
\end{equation}
Eq.~\ref{Chern2} is useful, as it allows one to treat finite disorder. A classic result \cite{BELLISSARD1994xj} states that the disorder average
$-2\pi i\left \langle  \sum_\alpha\langle 0,\alpha|P_\omega \big{[} [\hat{x}_1,P_\omega],[\hat{x}_2,P_\omega] \big{]}|0,\alpha\rangle \right \rangle_\omega,$ ($P_\omega$ = the projector onto the occupied states of $H_\omega$) is an integer if $E_F$ is in a region of localized EnS.  This integer can change its value \emph{only} if $E_F$ crosses a region of extended states, a property that allows one to map the delocalized spectrum. Moreover, if $H_0$ and $E_F$ are chosen such that $C$$\neq$$0$ at $W$=0, then moving in any direction in the $(E_F,W)$ plane from that initial point, $C$ will eventually become zero for one of the  reasons: 1) $P_\omega$=0 if $E_F$ is very negative, 2) $P_\omega$=1 if $E_F$ is very positive or 3) all the states localize if $W$ is too large. This implies the existence of a region of extended states surrounding the CI phase and explains the phase diagram of Fig.~\ref{PhaseDiagram}.

Unfortunately, Eq.~\ref{Chern2} only makes sense in the thermodynamic limit. We derive a finite size real-space formula for $C$ that converges exponentially fast to the thermodynamic limit. It does not involve twisted boundary conditions, which not only eliminates the problems associated with level crossings at $E_F$, but  allows us to compute $C$ for large systems and many disordered configurations. Additionally, our formula for $C$ requires only knowledge of the ground state.  

For clean CIs, $C$ is computed using a discretized Brillouin-zone: ${\bm k}_{\bm n}$=$n_1{\bm \Delta}_1$$+$$n_2 {\bm \Delta}_2$, $n_{1,2}$=$1,\ldots,N$, $|{\bm \Delta}_i|=\Delta=\frac{2\pi}{N}$. The partial derivatives $\partial_{k_{i}}\hat{P}$ are replaced by finite differences $\delta_i \hat{P}({\bm k}_{\bm n})$=$\sum_m c_m \hat{P}({\bm k}_{\bm n}+m{\bm \Delta}_i)$ and the integration by a Riemann sum. Because the integrand in Eq.~\ref{Chern1} is a periodic and analytic function, with an appropriate choice for the $\delta_i \hat{P}({\bm k}_{\bm n})$ approximation, the discretized formula converges exponentially fast to the continuum limit. To obtain our real-space representation, we note that the discretized $C$ formula can be written as $-i \mbox{Tr}\{P [\delta_1 P,\delta_2 P]\}$, where the trace is over the whole Bloch basis $\ket{\bm{k}_n \alpha}$ and $P$ is the full projector: $P=\sum_{\bm_n}  \ket{\bm{k}_{\bm n}\alpha } P_{\alpha \beta}(\bm{k}_{\bm n}) \bra{\bm{k}_{\bm n} \beta}$. Since the trace is invariant to a change of basis, we express this trace in the dual real-space basis $\ket{{\bm x},\alpha}$. The result is similar to that of Eq.~\ref{Chern2} but with the substitution ($c_m=-c_{-m}$):
\begin{equation}
\begin{array}{c}
-i[\hat{x}_i,P] \rightarrow \sum_m c_m e^{-im{\bm \Delta}_i\hat{{\bm x}}} Pe^{im{\bm \Delta}_i\hat{{\bm x}}}.
\end{array}
\end{equation}
If $c_m$'s are chosen so that: $x-\sum_{m=-N/2}^{N/2} c_m e^{i m x \Delta}=O(\Delta^N)$, the above substitution leads to exponentially small $O(\Delta^N)$ errors. Together with the localization of $P$, this leads to an exponentially fast converging formula.

For a clean CI ($\zeta$=$0.3i$), the formula gives $C$= 0.9999998/0.999999998 for a 30$\times$30/40$\times$40 lattice. The values at $W$=3 are shown in Fig.~\ref{chernandent}A. These calculations were performed for the 30$\times$30 and 40$\times$40 lattices, $10^3$ configurations and for seven $E_F$ values. The disorder averaged $C$ values (for the 40$\times$40 lattice) are 0.9999, 0.998, 0.85, 0.53, 0.17, 0.01 and 0.0001 for $E_F$=0, 0.5, 1.0, 1.5, 2.0, 2.5 and 3.0, respectively. These values indicate the existence of a delocalized spectral region between E=1.0 and 2.0, in good agreement with Fig. 1A. The topological character of the CI survives disorder.

\begin{figure}
\includegraphics[width=8cm, height=3.8cm]{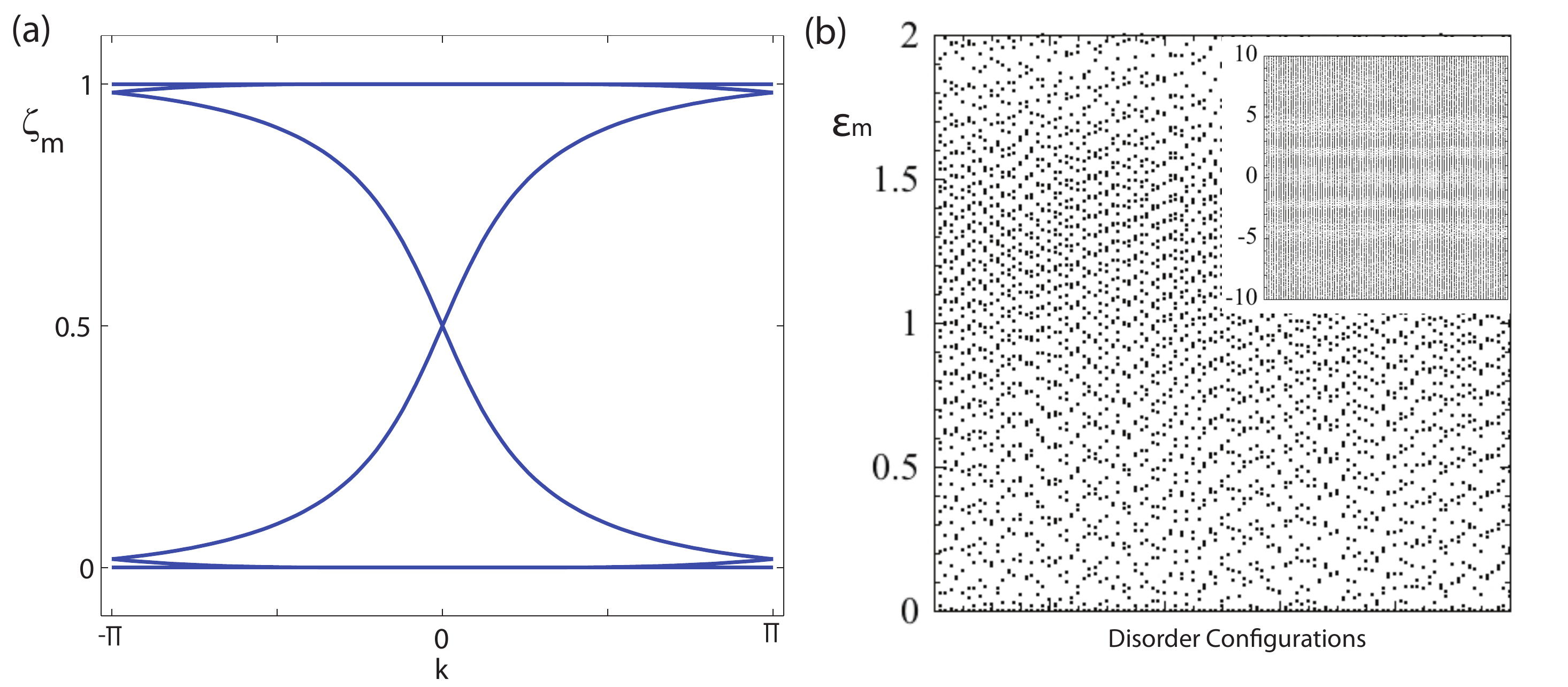}\\
\caption{a) Entanglement spectrum for a translationally invariant  CI plotted vs. momentum along the cut b) Level repulsion of the  ``edge" entanglement spectrum with disorder. Inset shows the full EtS.}\label{figureent}
\end{figure}

We now ask if the topological character of a CI has a clear signature in the EtS of the state.  We compute the reduced density matrix by cutting a torus shaped sample in two equal parts A and B, and then tracing out B.  The system is non-interacting, so we can use the single-particle EtS.  This involves  \cite{Peschel2004} finding the eigenvalues $\xi_m$ of the one-particle correlation function $C_{{\bm x}{\bm y}}$=$\langle c_{\bm x}^\dagger c_{\bm y} \rangle$, where ${\bm x}, {\bm y}$ are lattice sites of  \emph{the section that is not traced out}, and the expectation value is taken in the ground-state of the system.  The reduced density matrix can be decomposed in normal modes with energies related to $\xi_m$, $\rho_{A}$$\sim$$e^{- \sum_m  \epsilon_m a_m^\dagger a_m  }$ with $\epsilon_m$=$\frac{1}{2}\log\left(\frac{1 - \xi_m}{\xi_m}\right)$ where $a_k$ are normal mode operators.  The $\epsilon_m$'s are ``entanglement energies." In the clean limit, we perform translationally invariant cuts  and plot $\xi_m$'s as function of momentum along the cut, as in Fig \ref{figureent}(a). When $E_F$ is in the bulk gap, the $\xi_m$'s are primarily concentrated around $0, 1$ and have little dispersion. These are either bulk states deep in region A ($\xi_m$$\sim$1), or deep in region B ($\xi_m$$\sim$0). We call the difference between the levels at $\xi$=1 and the ones at $\xi$=0 the entanglement bulk gap.  For a trivial insulator, this is the whole story. For a nontrivial insulator as in  Fig \ref{figureent}(a), an entanglement mode localized on the cut crosses the entanglement bulk gap, much like an edge state in the EnS. As such, the EtS can differentiate between a topological and a trivial insulator, even though we are looking {\it only} at the \emph{bulk} ground-state wavefunction for a system \emph{without} boundaries.

We now add disorder.  Unfortunately, when $E_F$ is at half filing, the number of the $\xi_m$ levels that are away from $0$ or $1$ (i.e. the entanglement ``edge spectrum") is very small, about 15 for a 30$\times$30 lattice. As such, the large spacings between these levels render them in the clean regime of a random matrix. These ``entanglement edge" levels exhibit strong level repulsion as can be clearly seen by the naked eye in  Fig \ref{figureent}(b). The disorder mixing of these levels is small (unless we go to high disorder where the CI is destroyed) and the level-spacing variance, although small, differs from $0.178$. A computation with a larger 40$\times$40 lattice shows a decrease of the variance. This is  consistent with results in the almost clean limit of the EnS, where the disorder energy perturbation of each state is smaller than the mean level spacing \cite{Cuevas1998}. These levels exhibit level repulsion and are delocalized.

We now shift $E_F$ and recompute the EtS and the variance of its level spacings. The results are shown in Fig.~\ref{chernandent} for $W$=3. As we move $E_F$ towards the delocalized EnS, we notice that the level statistics of the EtS (in $\epsilon_m$) acquires an increasingly flat region of level spacings displaying a variance of $0.178$. As the $E_F$ is moved up from half filling, the EtS becomes more and more diffusive (i.e. departs from the clean limit of the half-filled finite size problem). Delocalized bulk levels (which are at large negative or positive $\epsilon$) start moving in. When $E_F$ sits right on top of the delocalized EnS, the \emph{whole} EtS becomes delocalized and has variance extremely close to the Wigner-Dyson surmise of $0.178$. The histograms of the level spacings collected from small windows at three widely spaced entanglement energies show well defined distributions matching closely the $P_{\mbox{\tiny{GUE}}}(s)$. The observation of this delocalization plateau in the EtS corresponding to the ground-state of the system filled up to the extended state energy is our main result. As $E_F$ is moved above the extended states and into the region of the trivial Anderson insulator, the entanglement spectrum starts to become localized, with the spectrum near the center starting first.  In contrast, for a trivial insulator, the EtS never has regions of level repulsion for any EtS of any ground-state. The calculations in Fig.~\ref{chernandent} were performed for 30$\times$30 and 40$\times$40 lattices which give similar results although with reduced noise for the latter. 

The behavior of the EtS can be understood by noticing that $C_{{\bm x}{\bm y}}$ is identical to the projection operator onto the occupied states, which can be regarded as a flat-band Hamiltonian. EtS is then \emph{identical} to the spectrum of this flat-band Hamiltonian \emph{with open} BC at the edge of the untraced region. The periodic BC flat band Hamiltonian has eigenvalues at $0$ and $1$, and  the open BC Hamiltonian will have most of its eigenvalues close to $0$ or $1$ as well. If the original system was an insulator, then the eigenvalues between $0$ and $1$ are separated by an entanglement gap, but if topologically nontrivial, the spectrum also has edge modes crossing the entanglement gap. This is an inescapable conclusion that requires no calculation. The projector onto occupied states satisfies the same symmetries as the original Hamiltonian. The projector Hamiltonian has long-range hoppings when $E_F$ is at the mobility edge, and we conjecture that this results in the flat, delocalized behavior of the \emph{full} EtS.  

In conclusion, the  EnS, EtS, and a new finite-size Chern number formula in the presence of strong disorder yield matching results and can be used to characterize the CI to Anderson-insulator transition. We found that \emph{all the levels} of the  EtS of a CI groundstate filled up to the \emph{ edge of the mobility gap} exhibit level repulsion consistent with the Wigner Dyson distribution (the many-body EtS matrices belong however to Wishart ensembles rather than Unitary ones). This delocalized  plateau in the entanglement spectrum could be used  to gain information about the many-body localization problem.

\emph{Acknowledgements} BAB was supported by Princeton  Startup Funds, Alfred P. Sloan Foundation and NSF DMR-095242, and thanks the Institute  for International Collaboration in Beijing, China for generous hosting. TLH  was supported in part by the NSF DMR 0758462 at the University of Illinois, and by the ICMT. EP acknowledges a support from the Research Corporation for Science Advancement.


%

\end{document}